\documentclass[a4paper,10pt]{article}
\usepackage{
times, amsmath,amssymb,epsfig,graphics,graphicx}

\newcommand{\be}{\begin{eqnarray}}
\newcommand{\ee}{\end{eqnarray}}

\begin{document}

\begin{center}
{\Large\bf Manifestation of Conformal Symmetry in the 
Light Flavor Baryon Sector }
\end{center}
\vspace{0.02cm}

\begin{center}
M.\ Kirchbach and C.\ B.\ Compean
\end{center}

\vspace{0.01cm}
\begin{center}
{\it Instituto de F\'{\i}sica}, \\
         {\it Universidad Aut\'onoma de San Luis Potos\'{\i},}\\
         {\it Av. Manuel Nava 6, San Luis Potos\'{\i}, S.L.P. 78290, M\'exico}
\end{center}

\vspace{0.01cm}

\begin{flushleft}
{\bf Abstract:}
On the AdS$_5$ cone, conformally compactified to
{\bf R}$^1\otimes S^3$, a quark-diquark model of light flavor 
baryons is developed.  The system on this manifold 
is described in terms of a scalar conformal equation, 
gauged  by  the field of a D3 brane whose
transversal dimensions have been conformally wrapped  
over {\bf R}$^1\otimes S^3$.
Such a system does not result exactly conformally invariant because
the gauge potential slightly modifies
the  conformal centrifugal barrier of the free geodesic motion.
Thanks to this, the model describes the correct mass ordering in the
$P_{11}-S_{11}$ pairs through the $N$ spectrum as a combined  
effect of conformal symmetry breaking, on the one side,
and a parity change of the diquark from a scalar at low masses,
to a pseudoscalar at higher masses, on the other.
We calculate the number of resonances with masses below 2500 MeV
needed for the completeness of the above scheme and find a total
of 32 ``missing'' nucleon and $\Delta$ states.
Their absence or presence in the respective spectra relates to the degree
to which conformal symmetry is realized in the lightest flavor sector. 
The scheme also allows for a prediction of
the dressing function of an effective 
instantaneous gluon propagator from the Fourier transform of 
the gauge potential. We find a dressing function that is finite in the 
infrared and tends to zero at infinity.
\end{flushleft}

\section{Introduction}

Applications of brane theory to hadron physics have
acquired considerable attention in recent times. The possibility for such 
applications arose in effect of the intriguing observation
\cite{Nicolai}-\cite{Mack} that the asymptotic horizon geometry
of the Dirichlet three-brane (D3) of the
IIB superstring in ten dimensions considered  on 
the $ AdS_{5}\otimes S^{5 }$ background,
admits a superalgebra that is identical to the super-conformal algebra
of the corresponding 4d  world-volume 
field theory  when gravity is decoupled.  
The D3 brane theory has $SU(2,2/4)$  as underlying
superconformal symmetry  whose bosonic isometry 
 $SU(2,2)\otimes SU(4)$, is  locally isomorphic to
$ SO(2,4)\times SO(6)$ \cite{Frr_Frnsdl}.
This  group happens to coincide with
the isometry group of the corresponding 
asymptotic horizon background, $AdS_5\times S^5$, 
a mathematical coincidence
that was suggestive of a duality between supergravity 
around the horizon background, on the one side,
and superconformal brane dynamics, on the other.
It has been conjectured by Maldacena \cite{Wilson} that  
zero temperature super Yang-Mills theory 
residing in the conformal $AdS_{5}$ boundary is likely
to capture some of the essential  features of high-temperature 
3d  QCD.  
Within this context, testing AdS/CFT reduces to
the calculation of observables within that very framework 
and their comparison to the corresponding 
Lattice results in 3d  QCD.
The duality
between D3-brane bulk supergravity and super Yang-Mills open string theories
on the conformal boundary  
of $AdS_{5}$ space-time  implies  that  each CFT state 
(among them the QCD states in the light flavor sector)
can be put in correspondence  to a state in the 
supergravity approximation to string theory
on $AdS_{5}\otimes S^5$ meaning that spectra 
in both theories should come out same.
As long as the isometry group of the conformal  
$AdS_{5}$  boundary  is the conformal group
$SO(2,4)$,  the requirement is that the  quantum states in the
theory on that  boundary  should populate $SO(2,4)$ 
unitary representations \cite{Seiberg}.
This is an essential restriction which 
strongly limits the number of theories respecting AdS$_5$/CFT$_4$
and is suggestive of the construction of quark models 
that respect global conformal invariance.
Conformal symmetry is independently 
to a good approximation a global symmetry of 
the QCD Lagrangian in the light-flavor sector, a reason 
for which one can expect  spectroscopic data 
on the light flavor baryons, 
the nucleon and the $\Delta$, to be especially
appropriate in examining the AdS/CFT concept.
Implementations of  AdS/CFT  to hadron physics have been 
pioneered in refs.~\cite{Br1} within the framework of  light-front 
QCD. There, the conformally invariant two-dimensional (2d) Hamiltonian,
$\nu(\nu+1)/\zeta^2 + \omega^2 \zeta ^2$,  with
$\zeta $ standing for the so called weighted impact variable, and
$\nu $ related to twist,  has been successfully
employed in the description of both the meson and light baryon sectors.
Further interesting applications of the holographic concept to
hadron physics can be found, among others, in refs.~\cite{dong}--
\cite{Nawa}.
We here instead implement conformal symmetry into a quark Hamiltonian
in three-dimensions  which explicitly contains the
ordinary angular momentum operator, a circumstance that favorably
facilitates identification  of predictions to data. We reveal possibility of 
designing such a model 
in making use of the  conformal map
of  ordinary Minkowski space, ${\mathcal M}=${\bf R}$^{1+3}$ 
onto  $S^1\otimes S^3$ \cite{Nicolai}, or, {\bf R}$^{1}\otimes S^{3} $    
\cite{Mack}.
According to \cite{Mack}, correlation functions of 
CFT on {\bf R}$^{1+3}$ can be analytically continued to the 
full Einstein universe, this because {\bf R}$^{1+3}$ 
can be conformally mapped on {\bf R}$^1\otimes S^3$.
The implication of this important observation is that
each CFT state on 
{\bf R}$^1\otimes S^3$ can be brought into unique correspondence with a 
state of the brane theory on $AdS_{5}\otimes S^5$. 
Consequences  on thermal states have been worked out in ref.~\cite{Horow}. 
We here examine consequences for the systematics
of the $N$ and $\Delta$ spectra.

\noindent
The contribution is organized as follows.
In the next section we highlight the procedure of
conformal compactification of Minkowski spacetime
along the line of ref.~\cite{Gibbons} and present the
conformal equation which will be applied to
a  quark-diquark (q-(qq)) model of light baryons in section 3.
There, we calculate the nucleon spectrum 
and the mean square charge radii 
of some of the excited states.
Section 4 is devoted to the  design of a dressing function of 
an (instantaneous) effective gluon 
propagator as a Fourier transform of the gauge potential on
$S^3$.
The contribution closes with brief conclusions.
\section{ Conformal compactification of the AdS$_5$ cone to 
$S^1\otimes S^3 \sim $R$^{1}\otimes S^3$ }

The description of the confinement phenomenon is one of the
major goals in the physics of hadrons. Confinement implies
exclusion of 
scattering states and favoring  discrete bound states alone.   
Putting systems on finite volumes
is a standard strategy of spectrum discretization \cite{barbon}.
There is a variety of geometries appropriate for  
preserving  the conformal symmetry by the resulting Hamiltonians,
the three dimensional sphere, $S^3$, being one of them 
\cite{Mack}, \cite{Gibbons}, \cite{barbon}.
It has been shown that  geometries containing $S^3$, such as the 
 $S^1\otimes S^3$  manifold,
can be  approached departing
directly from AdS$_5$ \cite{Nicolai},\cite{Gibbons1}, \cite{Gibbons}. 
The five dimensional manifold AdS$_5$ is  defined as a 
{\bf R}$^{2+4}$ subspace according to 
\begin{equation}
u^2+v^2 -x_1^2-x_2^2-x_3^2-x_4^2=\rho^2,
\end{equation}
where $\rho $ is a fixed parameter.  The boundary at infinity of this space
is identified with the $AdS_5$ cone
\begin{equation}
u^2+v^2 -x_1^2-x_2^2-x_3^2-x_4^2=0.
\end{equation}
Flat four-dimensional (4d) Minkowski space time can then be thought of as the 
intersection of the null hyperplane $v^2-x_4^2=0$ with the
$AdS_5$ cone, in which case 
\begin{equation}
v^2-x_4^2=u^2-x_1^2-x_2^2-x_3^2=0,
\label{LCone}
\end{equation}
and $u$, and ${\vec x}=$column$(x_1,x_2,x_3)$
in turn assume the r\'oles of time and position vector in
{\bf R}$^{1+3}$, respectively.
A quark model placed on the AdS$_5$ cone patterned after eq.~(\ref{LCone})
allows for a description in terms of light-front variables, a path taken by
refs.~\cite{Br1}. 
We here instead take the different view on
Minkowski space as emerging from an
AdS$_5$ cone patterned after
\begin{equation}
u^2+v^2=x_1^2+x_2^2+x_3^2+x_4^2=R^2, \quad R\not=0, 
\label{null_rays}
\end{equation}
in which case the cone  compactifies to 
${\mathcal M}^{\ast( 1+3)}=S^1\otimes S^3$.
The 
$S^1\otimes S^3$ manifold then describes  
the particular set of {\bf R}$^{2+4}$ null rays
associated with  eq.~(\ref{null_rays}).
As long as the isometry group,
$SO(2)\otimes SO(4)$,  of $S^1\otimes S^3$,
is a subgroup of the conformal group $SO(2,4)$ of 
regular (1+3) Minkowski space, ${\mathcal M}$, i.e.
$SO(2)\otimes SO(4)\subset SO(2,4)$, 
the above procedure  is referred to as
conformal compactification  of Minkowski space time. 
Now one can parametrize the $S^1\otimes S^3$ manifold by the four angles 
$\tau, \chi, \theta, \varphi$ in accordance with
\begin{eqnarray}
u+iv=Re^{i\tau }\, &\quad& x_1+ix_2=R\sin\chi \sin\theta e^{i\varphi},\quad
x_3=R\sin \chi \cos \theta, \nonumber\\
x_4^2+{\mathbf r}^2=R^2,&\quad&
 r= |{\mathbf r}|=R\sin\chi,
\label{RW_prmtrz}
\end{eqnarray}
where  $R$  is the $S^3$ hyper-radius. 
According to ref.~\cite{Gibbons},  at the microscopic scale
this map takes the
flat space Minkowski metric to the metric of
Einstein's  {\bf R}$^1\otimes S^3$ cylinder, 
\begin{equation}
ds^2=\Omega ^{-2}
(-{\mathrm d}\tau^2 +
{\mathrm d}\chi^2 +\sin^2\chi
({\mathrm  d}\theta^2 +\sin^2\theta 
{\mathrm d} \varphi ^2)),
\label{metric}
\end{equation}
with $\Omega$ being the conformal factor. In this way,
$S^1\otimes S^3\simeq ${\bf R}$^{1}\otimes S^3$ isomorphism  
is established

\subsection{Free geodesic motion on $S^3$ and the conformal free rigid rotor.}

Within the metric of eq.~(\ref{metric}), 
and the conformal factor being absorbed by the 
wave functions,  the following
conformally invariant massless scalar field equation 
has been found in ref.~\cite{Gibbons},
\begin{equation}
-\hbar^2 {\widehat \Box }\psi +\mu^2 \psi =0.
\label{conf_eq}
\end{equation}
Here, $\mu^2$ is a conformal constant \cite{Gibbons}, 
${\widehat \Box}$  stands for the angular part of the 
4d Laplace-Beltrami operator, which we here choose to express in terms of
${\mathcal K}^2$, and  ${\mathbf L}^2$,    
the operators of the squared
four- and three dimensional
angular momenta as,
 \begin{eqnarray}
{\widehat \Box} &=&  
-\frac{1 }{R^2}\frac{\partial ^2}{\partial \tau ^2}-
\frac{1}{R^2}{\mathcal K}^2,\quad \kappa=\frac{1}{R^2},\quad
-{\mathcal K}^2=  \left[\frac{1}{\sin^2\chi }
\frac{\partial }{\partial \chi}
\sin^2\chi \frac{\partial }{\partial \chi } -
\frac{{\mathbf  L}^2 (\theta ,\varphi ) }{\sin^2 \chi }\right].
\label{lpls_5}
\end{eqnarray}
Furthermore, $\chi \in [0,\pi ]$ is the second polar angle on $S^3$,
 while $\kappa $ denotes the curvature.
The ${\mathcal K}^2$ eigenstates, $|Klm>$, are well known to
belong to irreducible  $SO(4)$ representations of the type  
$\left(\frac{K}{2},\frac{K}{2} \right)$, and the quantum numbers, 
$K$, $l$, and $m$ define the eigenvalues of the respective  
four--, three-- and two--dimensional angular momentum operators
upon the  states \cite{Kim_Noz}. The ${\mathcal K}^2$ eigenvalues, 
$\lambda _K$, upon $|Klm>$, are known too and given by, 
\begin{eqnarray}
{\mathcal K}^2 \vert K l m \rangle = {\lambda }_K
\vert K l m \rangle,
&& {\lambda  }_K=K(K+2), \quad
\vert Klm\rangle \in \left( \frac{K}{2},\frac{K}{2} \right),
\quad  \quad K=,0,1,2,..., \infty,
\nonumber\\
{\mathbf L}^2 |Klm\rangle =l(l+1)|Klm\rangle ,
&\quad&  L_z|Klm>=m|Klm\rangle, \quad
l=0,1,...K, \quad  m=-l,..0,...l.
\label{Casimir_O4}
\end{eqnarray}
The infinite series of  solutions of 
eq.~(\ref{Casimir_O4}) constitute 
an $\infty$d unitary representation of the conformal group that has been
built up from the
eigenstates of its little group, $SO(4)_{K}$ \cite{Frr_Frnsdl},
\cite{Wybourne}. This irrep will be frequently 
termed to as a ``conformal band''.
Therefore, the conformal symmetry aspect of AdS/CFT   
is adequately captured by the ${\mathcal K}^2$ eigenvalue 
problem, which in this fashion qualifies  
as a suitable departure point toward the description of  
conformal excitation modes in two-body systems.
Independently, conformal symmetry is also to a good approximation global
symmetry of the QCD Lagrangian in the light flavor sector, one more reason
why employing the conformally invariant AdS/CFT scenario from above
in modeling $N$ and $\Delta $ excitations should be of interest.
Upon factorizing the $\tau $ dependence of the solution to
eq.~(\ref{conf_eq}) as 
$\exp {\left( i\frac{
{\mathcal E}}
{\hbar\sqrt{\kappa } } \tau \right)
}$,  one arrives at the following
angular equation  on $S^3$,
\begin{eqnarray}
{\Big[} {\hbar^2}\kappa {\mathcal K}^2 
&-& {\mathcal E}^2+\mu^2  {\Big]} 
{\mathcal S}  (\chi)=0.
\label{chi_eq}
\end{eqnarray}
Though in angular variables, the conformal eq.~(\ref{chi_eq})
has the algebraic form of a Klein-Gordon equation which provides a 
relativistic description of the free geodesic motion on $S^3$
in terms of  the eigenvalue problem of the squared 4d 
angular momentum.
The spectrum of eq.~(\ref{chi_eq}) reads,
\begin{equation}
{\mathcal E}^2_K -\mu^2 =\kappa {\hbar^2}K\left(K+2\right).
\label{4D_rotator}
\end{equation}
and represents no more but what one could term to as   
the spectrum of the ``conformal free rigid  rotor''.
Stated differently, 
one can say that the ${\mathcal K}^2$ eigenvalue problem
has $SO(4)$ as potential algebra, and 
the conformal group as dynamical symmetry. \\

\subsection{The conformal interacting rigid rotor.}

The next step is introducing  interaction on $S^3$ in such a manner as
to respect the conformal symmetry of the spectrum of the free geodesic motion. 
For this purpose,  use can be made of an observation reported
in ref.~\cite{Hrwtz}. There, a broad class of
exact string solutions have been constructed
by wrapping transversal dimensions of
fundamental strings over curved spaces and solving the 
corresponding curved space Laplace-Beltrami equations for harmonic functions.
Specifically on $S^3$ it has been shown that a conformal string field,
$K(\chi)$, (in the notations of ref.~\cite{Hrwtz}) is 
obtained from the metric,
\begin{equation}
ds^2={\mathrm d}u{\mathrm d}v +  K(\chi) {\mathrm d}u^2 + {\mathrm d}\chi^2
+\sin^2\chi{\mathrm d}\theta^2 +\sin^2\theta {\mathrm d}\varphi^2,
\label{Horus1}
\end{equation}
with $K(\chi)$ being a harmonic function on $S^3$
\begin{equation}
{\widehat \Box} K(\chi) = 0, \quad K(\chi )= {\bar a}+m\cot\chi .
\label{Horus2}
\end{equation}
This  important observation  qualifies 
the cotangent function 
as a conformal potential on $S^3$. 
A further and independent motivation in favor of
employing the cotangent potential in quark models 
is provided by the observation \cite{CK_07} that
the lowest terms in  its Taylor expansion
coincide with a Coulombic+linear (Cornell) potential. This is best illustrated
by the simplistic  $\chi=\frac{r}{R}\pi $ parametrization
(commonly used in super-symmetric quantum mechanics) for which 
\begin{equation}
-\cot \frac{r}{R}\pi =
-\frac{d}{r}+\frac{1}{3}\frac{r}{d} +
\frac{r^3}{45d^3}+
\frac{2r ^5}{945 d^3}+..., \quad 
\mbox{with}\quad  d=\frac{R}{\pi },
\quad m>0,
\label{crnl}
\end{equation}
holds valid.
The Cornell potential \cite{Cornell} has been
predicted by lattice QCD simulations \cite{Lattice}, on the one side,
and  has been also independently confirmed within the AdS/CFT context where
it emerges as a soft-wall Wilson loop potential
\cite{Wilson},\cite{Wen}. Corrections to the Cornell potential
that account for non-perturbative processes, 
have been calculated  within the topological approach  
\cite{Brambilla}.
Within the light of this discussion, the finite range
of the cotangent potential can be viewed as a particular phenomenological 
parametrization of the non-perturbative corrections to 
the Coulombic+linear potential embedded in it.
Therefore, the cotangent potential on $S^3$, 
besides being congruent with 
AdS$_5$/CFT$_4$,  also correctly captures the dynamical aspects of 
QCD, a twofold advantage that  makes it attractive to 
applications in  hadron spectroscopy.

\noindent
We shall introduce this very potential  as a gauge
interaction in eq.~(\ref{lpls_5})
by means of the replacement, 
\begin{eqnarray}
i\hbar \sqrt{\kappa} \frac{\partial }{\partial \tau} &\to &
i\hbar \sqrt{ \kappa} \frac{\partial }{\partial \tau} + 
2G\sqrt{\kappa }\cot \chi -{\bar a},\nonumber\\
\label{intr_intr}
\end{eqnarray}
where we parametrized $m$ in eq.~(\ref{Horus2}) as $m=-2G\sqrt{\kappa}$.
Upon factorizing the $\tau $ dependence of the total wave function
as $\exp \left(-i \frac{E}{\hbar \sqrt{\kappa}}\tau\right)  $, and after 
some algebraic manipulations, 
the interacting Klein-Gordon 
equation can be cast into the following form,
\begin{eqnarray}
{\Big(}-\kappa {\hbar^2}\frac{\mbox{d}^2 }{\mbox{d}\chi ^2} 
+ U_l(\kappa, \chi )
 & - &2G \sqrt{\kappa}(2E-2{\bar a})  \cot\chi
-(2G\sqrt{\kappa})^2 \csc^2\chi
 {\Big)}\Psi (\chi ) \nonumber\\
&=&
\lbrack (E-{\bar a})^2-\mu^2 
- \left( {\bar a}^2+ (2G\sqrt{\kappa})^2\right)\rbrack \Psi (\chi) . 
\label{chi_eq_intr}
\end{eqnarray}
The second  term on the l.h.s of this equation,
\begin{equation}
U_l(\chi, \kappa)= \hbar^2\kappa l(l+1)\csc^2\chi,
\label{s3_brr}
\end{equation}
is the centrifugal barrier of the free geodesic motion on $S^3$.
Indeed, introducing the new constant,
\begin{eqnarray}
\alpha (l) &=&
-\frac{1}{2} +\sqrt{
\left( l+\frac{1}{2}\right)^2
-\frac{(2G)^2}{\hbar^2}}
= l+\Delta \, l,\quad \Delta\, l\approx  
-\frac{1}{2}\frac{(2G)^2}{\hbar^2}\frac{1}
{l+\frac{1}{2}},
\label{new_a}
\end{eqnarray}
allows to rewrite eq.~(\ref{chi_eq_intr}) equivalently to,
\begin{eqnarray}
 \lbrack  -\kappa {\hbar^2}\frac{\mbox{d}^2 }{\mbox{d}\chi ^2} 
+{\mathcal V}(\chi )\rbrack \Psi (\chi )&=&
((E-{\bar a})^2- c_0)\Psi (\chi)  \nonumber\\
{\mathcal V}(\chi )=
-2\beta \cot\chi  + {\overline U}_l(\chi, \kappa ) , &\quad &
{\overline U}_l(\chi, \kappa )=\hbar^2 \kappa \alpha (l) 
(\alpha (l) +1) 
\csc^2\chi, \nonumber\\ 
\beta = 2G \sqrt{\kappa}(E-{\bar a}),&\quad&
c_0= \mu^2 -\hbar^2\kappa + {\bar a}^2 + (2G\sqrt{\kappa})^2.
\label{chi_ctgnt}
\end{eqnarray}
The positive sign in front of the square root in eq.~(\ref{new_a}) 
ensures that ${\overline U}_l(\chi,\kappa)$
approaches  the $S^3$ centrifugal barrier of the free geodesic motion
in the $G\to 0$ limit, 
${\overline U}_l(\chi,\kappa )
\stackrel{G\to 0 }{\longrightarrow}U_l(\chi, \kappa).$
Upon a suitable variable change, differential equations of the type
in (\ref{chi_ctgnt}) have been shown \cite{raposo}
to reduce to one of the
forms of the hyper-geometric
equation whose solutions can be  expressed in closed form 
in terms of the non-classical Romanovski polynomials
(here denoted by $R_n^{(a,b)}(\cot \chi )$) .
The resulting explicit formula
for the wave functions then emerges as,
\begin{eqnarray}
\Psi_{Kl} (\chi)&=&
N_{Kl } e^{-a {\chi}}(\sin \chi )^{K+1+ \Delta l}
R_n^{(a,b)}(\cot \chi),
\label{psil} \\
a= \frac{2G(E-{\bar a})}{\sqrt{\kappa}\hbar^2 (K+1 +\Delta l)}, &\quad&
b=-(K+1 +\Delta l ), \quad K=n+l,
\label{constants}
\end{eqnarray}
where $N_{Kl}$  are normalization constants.
Correspondingly, the algebraic equation for the
energy  takes the form,
\begin{eqnarray}
(E-{\bar a})^2 -c_0&=&{\hbar^2}\kappa (K+1 +\Delta l)^2 
-\frac{4G^2\kappa(E-{\bar a})^2}{\hbar^2}\frac{1}{(K+1+\Delta l)^2},
\quad l\in [0, K].
\label{mss_frml}
\end{eqnarray}
In the interacting case, the principal quantum number, 
$(K+1 +\Delta l)$, in eq.~(\ref{mss_frml}) 
can be at most close to integer,
thus relaxing the degeneracy of states within the $(K/2,K/2)$
multiplet in eq.~(\ref{Casimir_O4}).
Nonetheless, it will still be possible to classify a light baryon
spectrum in terms of  the $SO(4)$ states $|Klm>$,
and this very spectrum will  still fall into an 
$\infty$d unitary representation of
the conformal group. This because, as we shall see below,
the $\Delta l $ contributions will have a detectable effect only  
on the masses of states with $l=0$ and will be
helpful  in removing the degeneracy between
$P_{2I,1}$--$S_{2I, 1}$ resonances and
reproducing their correct mass orderings,
while leaving all the other excitations  
practically degenerate. 
In effect, the relativistic  framework of the Klein-Gordon equation
(\ref{chi_eq_intr}) with a
cotangent gauge potential provides the 
intriguing possibility of having  conformal bands whose states are 
not necessarily perfectly degenerate
though they still keep spreading around $SO(4)$ levels. 
This is quite an interesting situation and
contrasts the case of the linear in
the energy Schr\"odinger equation which keeps
respecting in the interacting case the degeneracies of the free
geodesic motion \cite{CK_09}.

\section{Quark-diquark model on $S^3$ and the $N$ spectrum  }
From now onward  we assume dominance of quark-diquark configurations
in the internal nucleon and  $\Delta$ structures and apply 
eq.~(\ref{chi_eq_intr}) to  the description of the relative motion
of these two bodies. In considering the diquark as spinless,
the total spin, $J$,  of the resonance is then obtained
through coupling the spin-1/2 of the quark to the $q-(qq)$ 
relative angular momentum, $l$, according to eq.~(\ref{Casimir_O4}).
For the time being, and because of the absence of spin-flavor 
interactions in the
wave equation under consideration,
we shall factorize the light flavor quantum number
(actually isospin).
In due course we shall see that the spectra reported so far
do not contradict  the above assumptions. 
As a working hypothesis, to be tested by comparison with data,
the nucleon diquark has been set as a scalar  
for the ground state and the low lying excitation with masses below 
~1600 MeV, and as a pseudo-scalar  above. 
The $\Delta$ diquark has been set as an axial vector in the 
ground state and as a pseudo-scalar above.
The pseudo-scalar  diquark 
is a $P$ wave axial-vector  of zero total angular momentum and
emerges from coupling the axial spin vector, ${\vec S}={\vec 1}^+$, 
to an internal vectorial, $1^-$, excitation of the diquark according to
$\lbrack {\vec 1}^+\otimes {\vec  1}^-\rbrack ^{0^-}$. 
The quark-diquark system is described by means of 
eq.~(\ref{chi_ctgnt}).
The energies $E$ in eq.~(\ref{mss_frml})  
will be subsequently re-denoted by $M$ and
given interpretation of resonance masses read off
from the respective ground state nucleon and $\Delta$ masses.
In this fashion, the spatial part of the
total baryon wave function has been designed to account for
conformal symmetry in accord with AdS/CFT on the one side, and
with the (approximate) conformal symmetry of the QCD Lagrangian in the
light flavor sector, on the other.  
Next we shall compare the outcome of such a model
with data on $N$ and $\Delta$ resonances.\\

\noindent
\underline{The nucleon spectrum.}
The spectrum of the nucleon continues being under debate despite 
the long history of the respective studies (see refs.~\cite{Lee}, 
\cite{Afonin} for recent reviews).
Yet, unprejudiced inspection of the data reported by the Particle
Data Group \cite{PART} reveals systematic and hardly to overlook
grouping of the
excited states of the baryons of the best coverage, the
nucleon $(N)$ and the $\Delta (1232)$. 
Take as a prominent example the seven $\Delta $ resonances 
$S_{31}(1900)$, $P_{31}(1910)$, $P_{33}(1920)$,$D_{33}(1940)$,
$F_{35}(1905)$,   $D_{35}(1930)$, and $F_{37}(1950)$, which are
squeezed in the narrow mass band between 1900 MeV to 1950 MeV
and which, given the limitations  of data accuracies,
are practically mass degenerate.
This group of resonances consists of 3 parity dyads with spins ranging
from $\frac{1}{2}^\pm$ to $\frac{5}{2}^\pm$ and of a single parity state
of maximal spin, $J^\pi =\frac{7}{2}^+$ and its quantum numbers fit
into the $K=3$ multiplet in eq.~(\ref{Casimir_O4}) .
In the nucleon spectrum, one finds the 
$N(1440)$--$N(1535)$--$N(1520)$ triplet, which 
would match quantum numbers of $K=1$ in eq.~(\ref{Casimir_O4}).
Patterns of similar type do not restrict to these two examples 
alone but extend to the spectrum  of any of the lightest 
flavor baryons, no matter $N$, or $\Delta$.
It verifies directly by inspection  that the mass-splittings 
inside the above groups are notably smaller than the splittings 
between the averaged cluster masses. In  this fashion  the $\Delta $ spectrum
reveals a level structure.
The $N$ excitations follow quite same patterns though  appear shifted 
downward in mass by about 200 MeV relative to the $\Delta$  excitations.
Attention to this clustering phenomenon in light baryon spectra
has been drawn  in ref.~\cite{MK_01}.
In assuming equality of the quantum numbers of the $N$ and $\Delta$ 
excitations and comparing both spectra, allows to pin down the states
missing for the completeness of the scheme.

\noindent
The above clustering phenomenon
finds a quantitative explanation in terms of 
the spectrum of 
eqs.~(\ref{chi_ctgnt}),~(\ref{mss_frml}).(\ref{Casimir_O4})
under the assumption of  $q-(qq)^{0^\pm }$
as a dominant  configuration of internal baryon structure.
The parity, $\pi$,  of the di-quark, scalar, or, pseudoscalar, is fixed
by matching parity, ${P_{\mbox{\footnotesize max}}}$,  of the highest spin,
$J=\left(K+\frac{1}{2}\right)$,
of the multiplets in eq.~(\ref{Casimir_O4}),
the parity simplex, $J^{P_{\mbox{\footnotesize max}}}=
\left(K+\frac{1}{2}\right)^{P_{\mbox{\footnotesize max}}} $,
as $\pi =(-1)^K {P_{\mbox{\footnotesize max}}}$.\\

\noindent
\underline{Missing resonance predictions.}
The  comparison of the spectrum in eq.~(\ref{mss_frml})
to data on nucleon resonances \cite{PART} 
is presented in the figure. 
On the nucleon side the highest spins , 
$F_{17}$ from the $K=3$,  and $H_{31, 11}$ from 
$K=5$  are ``missing'' . The K=2 level is  completely 
``missing''. 
Compared to our previous work \cite{CK_07}, the
fit places the N(1900), N(1990), and N(2000) resonances at the
lower $K=4$ level and leaves instead $ P_{11},
F_{15},$ and $ F_{17}$ in K=5 unoccupied.
The number of ``missing'' nucleonic states is sixteen.
The $\Delta $ spectrum, not presented here, follows similar patterns.
We predict a total of 32  resonances ``missing'' from the $N$ and
$\Delta$ conformal bands.\\

\noindent
\underline{Mass ordering in $P_{2I,1}$--$S_{2I,1}$ pairs.}
The model predicts the correct mass ordering of the 
$P_{11}-S_{11}$ states through
the spectrum. 
Within the framework of the present study the numerical value of
the splitting between such states is entirely due to the
gauged  centrifugal barrier, ${\overline U}_l(\chi, \kappa )$,
defined in eqs.~(\ref{chi_ctgnt}),
which prescribes that $l=1$ states will  appear higher in mass than 
those with $l=0$. The ordering, $P_{2I,1}$--$S_{2I,1}$ versus
$S_{2I,1}$--$P_{2I,1}$ depends on the parity of the diquark.
When the diquark is a scalar, spin-$\frac{1}{2}^+$ and 
spin-$\frac{1}{2}^-$ in turn refer to 
zero and unit underlying angular momenta and 
are associated with $P_{2I,1}$, and  $S_{2I,1}$
states. In case of a pseudoscalar diquark, the situation reverses.
This is the reason for which at the scale of  1500 MeV,  
where the diquark is a scalar,
the measured  $P_{11}(1440)$ state appears  lower in mass
than its $S_{11}(1535)$ neighbor.  From
the 1700 MeV level onward, the parity of the diquark changes
to pseudoscalar, and it is $S_{11}$ that is built on top of a 
zero angular momentum.
Consequently,  $S_{11}$ states with masses above 1600 MeV
appear systematically at lower masses
than their nearest  $P_{11}$ neighbors. Examples are the
$S_{11}(1650)-P_{11}(1710)$, and $S_{11}(2090)-P_{11}(2100)$ pairs.
Recall that originally the suggestion of the parity change of the
diquark  was made with the purpose
of  matching parity of the highest spins in the fermionic multiplets
in eq.~(\ref{Casimir_O4}).  Therefore,
the reverse mass ordering in the $S_{11}-P_{11}$ pairs 
above 1600 MeV relative to the $P_{11}(1440)-S_{11}(1535)$ 
splitting, provides an independent argument
in favor of the change of parity of the diquark from 
scalar to pseudoscalar.\\

\noindent\underline{
Electric charge form-factors} are the simplest physical observables to 
calculate, and the corner stone of any spectroscopic model.
They reduce to the Fourier transform of the charge density, proportional
to $|\Psi_{Kl}(\chi )|^2$ in our case.  
The extraction of the mean square charge radius,
$<{\mathbf { r}}^2>$, from the form-factor
is then standard and calculated as the slope at origin.
We here wish to explore importance of relativistic corrections
and choose as illustrative examples the mean square charge radii of the
Nucleon, the $P_{11}(1440)$ and $S_{11}(1535)$ resonances.
The  wave functions 
entering the calculation are obtained from eq.~(\ref{constants}).
With the potential parameters fitted to the spectra, we find
$
\sqrt{<{\mathbf r}^2>_p}=0.664 \,\, \mbox{ fm},
$
to be compared to the experimental value of
$\sqrt{<{\mathbf { r}}^2>_p}=0.8750 \pm 0.008$ fm reported by
\cite{PART}.
For the Roper and the $S_{11}(1535)$ resonances, same observable is calculated 
as
$\sqrt{<{\mathbf { r}}^2>_{\mbox{\footnotesize Roper}}}=
0.8484 \,\, \mbox{ fm}, \quad \sqrt{<{\mathbf { r}}^2>_{
\mbox{
\footnotesize S}_{11}
}
}= 0.8754
 \,\, \mbox{ fm}.$ 
We obtain the mean square charge radius of the proton-like Roper resonance 
enhanced  by a bit less than 30$\%$  over the proton charge radius. 
Nonetheless, compared to the proton,
the form factor of the  $P_{11}(1440)$  takes smaller
values. This because our predicted $P_{11}(1440)$ charge density
appears slightly arced at origin.  
An enhancement, though smaller (10$\%$) has
been found by Nagata and Hosaka in ref.~\cite{Nagata}.
One of the  differences between the model by Nagata-Hosaka 
and the present model is that while in the former  both the 
scalar and axial vector diquarks are weighted by non-zero form factors,
we here  weight them by the extremal 
1 and zero values, respectively. This for the sake of staying as 
close to conformal symmetry as possible. In the present model, it is
the curvature parameter that seems to account for some of the effects
governed by the diquark form factors in flat space quark models.
Curvature as a phenomenological tool suited in mimicking  
complicated  many-body effects 
is known to be useful in the description of such complicated many-body
problems as Brownian motion, 
plasma correlations, instanton physics etc. \cite{Brown}, \cite{Nawa}. 
A reason for which  the replacement
of the complicated many-body problem of baryon structure 
(the genuine baryon wave function contains next to 
(q-(qq) also 3q, 3q($\bar q$ q)$^n$,3q(g$^n$) etc. configurations) by the
simple q-qq problem on $S^3$ turns out to be a useful approximation to reality
is that curvature, in combination with the conformal gauge 
potential, account reasonably for the omitted  many body effects.

\section{Dressing function for the gluon propagator}
The gluon propagator, $G_{\mu\nu}^{ab}(q^2)$, in the Landau gauge is
defined as
\begin{equation}
G^{ab}_{\mu\nu} =-i
\lbrack 
(
g_{\mu\nu}-
\frac{q_\mu q_\nu }{q^2}
)
\frac{G(q^2)}{q^2} \rbrack\delta^{ab}, \quad
 G(q^2) =\left(1+i\frac{m (q)}{q^2}\right)^{-1}.
\label{landau_gauge}
\end{equation}
In this gauge it is known to be
transverse in the Lorenz indices, and 
 $m (q)$ stands for the gluon self energy.
The function $G(q^2)$ is referred to as the 
gluon dressing function and can be obtained from solving Schwinger-Dyson
equations. In so doing, an especially  simple expression 
for $G(q^2)$ has been reported in \cite{Fedosenko} as
$G(q^2)\sim q^2$,
meaning finiteness of the gluon propagator in the infrared.
The propagator approaches zero in the ultraviolet.
More complicated expressions have been calculated in \cite{Aguilar}.
Independently, a finite gluon propagator in the infrared 
has also been  calculated recently 
in lattice QCD \cite{tereza}.
In view of these properties of the gluon propagator, it is of interest to
calculate the gluon dressing function from the {\bf R}$^{1}\otimes S^3$
quark model.  To do so we apply the Born approximation to
$E-V=\sqrt{{\mathbf p}^2 +\mu^2}$
and calculate the instantaneous ($q_0=0$)
gluon propagator in parallel to 
our recent work \cite{CK_09}
as a Fourier transform of the $\cot \chi(r) $ potential on
$S^3$ employing the integration volume in eq.~(\ref{RW_prmtrz}).
\begin{figure}
\resizebox{0.80\textwidth}{5.5cm}
{\includegraphics{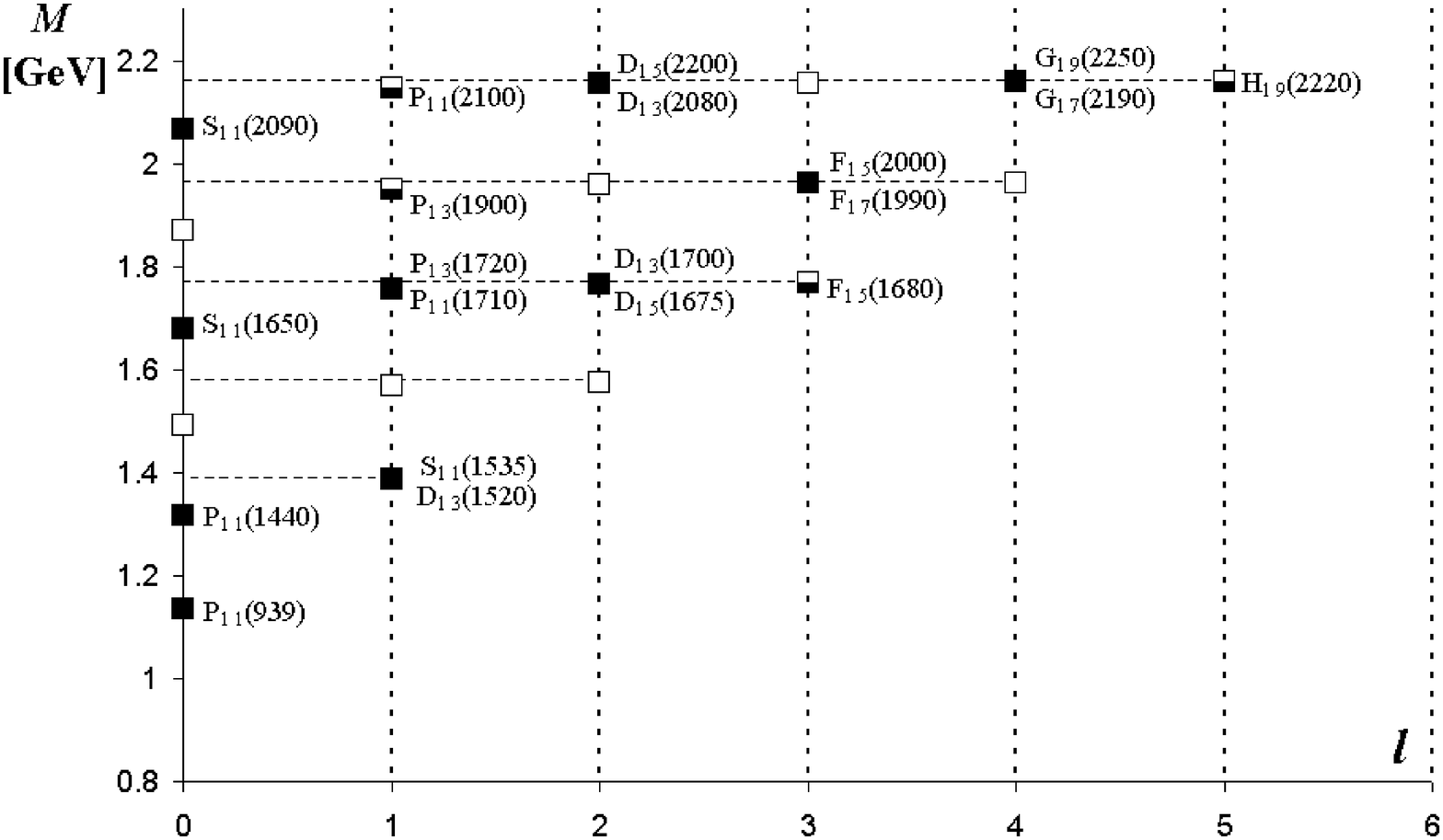}}
\caption{
Assignments on the $l/M$ grid 
of the reported \cite{PART} $N$ excitations to
the {\bf R$^1\otimes S^3$} spectrum in eq.~(\ref{mss_frml}).
The distribution of the reported (exp) resonances over the
predicted (th) states has been obtained from running a
four parameter least mean square data fit, i.e. 
minimizing
$\sigma=\sqrt{\frac{1}{N} \Sigma_{i=1}^{i=N} 
\left( M^{(i)}_{\mbox{\footnotesize th}}\, 
-M^{(i)}_{\mbox{\footnotesize exp }}\, \right)^2 }$.
The sum includes all the reported resonances.
The best fit that also predicts a reasonable value for
the proton mean square electric charge radius, is
$\sigma_{
\mbox{\footnotesize min}}=0.0855$ GeV 
has been obtained for
the following  potential parameters: 
$G= 0.0493$ GeV$\cdot $fm, $R=0.9814$ fm, and
$\mu =0.3213$ GeV, and ${\bar a}=0.932$ GeV..
The excitations, $E$, set to masses, 
have been read off from the  nucleon  
mass. Full and empty squares mark 
measured and ``missing'' resonances, respectively.}
\end{figure}
In Cartesian coordinates the $\cot \chi(r)$ term equals 
$\frac{x_4}{|{\mathbf r}|}$, and stands in fact for two potentials
distinct by a sign and describing interactions on 
the respective Northern, and Southern hemi-spheres.
Correspondingly, their respective  Fourier transforms 
to momentum space  become
\begin{eqnarray}
4\pi \Pi
( |\mathbf{q}| ) =\frac{(4\mu G\sqrt{\kappa})}{\hbar ^2} 
\int_0^\infty d|x| |x|^3 \delta (|x| -R)\int_0^{2\pi} d \varphi 
\int_0^\pi 
d\theta \sin\theta 
\int_{0/\frac{\pi}{2}}^{\frac{\pi}{2}/\pi } d \chi \sin^2\chi 
e^{i|\mathbf{ q}|\frac{\sin\chi}{\sqrt{\kappa}}|\cos\theta}
\cot \chi ,
\label{b2}
\end{eqnarray}
where the $\delta (|x| -R)$ function restricts {\bf R}$^4$ to $S^3$.
It is the requirement on invertability of this 
transformation, addressed here for the first time, 
that demands for distinguishing between  momentum 
space potentials on the Northern and Southern
hemispheres of $S^3$.
The first potential goes with $\chi \in {\Big[}0,\frac{\pi}{2}{\Big ]}$,  
corresponds to a positive $x_4$, and  describes an increasing 
$|{\mathbf r}|\in [0, R]$. The second one
refers
to $\chi \in {\Big[}\frac{\pi}{2},\pi {\Big ]}$, a negative $x_4$, 
and describes a decreasing $|{\mathbf r}|\in [R,0]$. 
To prove the invertability of the integral transform, 
 it is instructive to cast  the transformation integral
in eq.~(\ref{b2}) to the following equivalent  form,
\begin{eqnarray}
 4\pi \frac{\Pi (|{\mathbf q}|)}{\frac{(4\mu G\sqrt{\kappa}) }
{\hbar ^2} }&=&\pm 
\int_{0}^{R}{\mathrm d}r\frac{r^2}{\sqrt{R^2-r^2}} 
\frac{\sqrt{R^2 -r^2}}{r}
\int_0^{2\pi} d \varphi 
\int_0^\pi 
d\theta \sin\theta e^{i{\vec q}\cdot {\vec r}}
= \pm 4\pi  \frac{1-\cos q R}{q^2},
\label{uw_1}
\end{eqnarray}
where we used the parametrization in eq.~(\ref{RW_prmtrz}).
Applying now the inverse transformation on eq.~(\ref{uw_1}), 
results in
\begin{eqnarray}
&&\pm \frac{1}{ (2\pi )^3}\int _0^\infty
{\mathrm d}q \, q^2\int_0^{2\pi }{\mathrm d}\varphi
\int_0^\pi {\mathrm d}\theta \sin \theta  
\frac{4\pi (1-\cos q R) }{q^2}
\sqrt{R^2-r^2}e^{-i{\vec q}\cdot {\vec r}}
\nonumber\\
&=&\pm\frac{2}{\pi }
\frac{
\sqrt{R^2-r^2}
}{r} \{ 
\begin{array}{cc}
\frac{\pi}{2}, \qquad r<R,\\
\frac{\pi}{4}, \qquad r=R,\\
0, \qquad  r>R,
\end{array}
\end{eqnarray}
which proofs the invertability.
We here for concreteness pick up the  Northern 
hemisphere potential  and cast it in the more compact form,
\begin{equation}
\Pi(|{\mathbf q}|)= 
c \frac{2\sin^2 
\frac{|{\mathbf q} |}{2
\hbar \sqrt{\kappa}}}{\left(\frac{|{\mathbf q}|}
{\hbar\sqrt{\kappa }}\right)^2}, \quad c= \frac{4G\mu }{\hbar^2\kappa  }.
\label{prop_we}
\end{equation}
It is increasing  in the infrared,
finite at origin, and approaches asymptotically  the Coulomb 
propagator in the ultraviolet.
In the notations of eq.~(\ref{landau_gauge}) our result takes the form
\begin{equation}
\frac{G({\mathbf q} ^2)}{{\mathbf q} ^2}=
c \frac{2\sin^2 
\frac{{\mathbf q} }{2}}{\left({\mathbf q}\right)^2}, 
\end{equation}
for a dimensionless  ${\mathbf q}$  
measured in units of $\hbar\sqrt{\kappa}$. 
Stated differently,
\begin{equation}
G( \mathbf{q} ^2)=2c\sin^2 \frac{|{\mathbf q} |}{2}=
c(1-\cos |{\mathbf q} |)=c
(\frac{{\mathbf q}^2}{2!}-\frac{{\mathbf q}^4}{ 4!}+
\frac{{\mathbf q}^6}{6!}-...),
\end{equation}
 has been found. Therefore,
quark physics in {\bf R}$^{1}\otimes S^3$ also predicts 
a finite gluon dressing function in the infrared which  
approaches zero in the ultraviolet. 
Such a type of behavior has been observed, for example,
in the description of confinement phenomena  \cite{tereza}.

\section{Conclusions}
We examined consequences
of conformal symmetry in gravity-gauge duality
on spectroscopic data  on the lightest baryons,
the nucleon and the $\Delta (1232)$. The AdS$_5$/CFT$_4$ concept has been
implemented by a quark-diquark model on the AdS$_5$ cone, conformally
compactified to {\bf R}$^1\otimes S^3$. The description of the
$q-(qq)$ system on this compact manifold has been executed in terms
of the scalar conformal equation there, gauged by the field of
a D3 brane whose transverse dimensions have been conformally wrapped   
over $S^3$. The spectrum of such a two-body system falls as a whole
into a $\infty$d unitary representation of the conformal group, 
whose levels are irreps of $SO(4)$, the maximal compact group of
$SO(2,4)$. For such $SO(2,4)$ irreps, the notion of  
 ``conformal bands'' has been used. 
We observed that all nucleon and $\Delta$ resonances listed so far by the 
Particle Data Group distribute each fairly well over the  first five levels of
a respective conformal band. 
We did not exclude none of  the states from the analyzes.
We predicted a  total of 32 ``missing'' resonances
needed for the completeness of the conformal nucleon and 
$\Delta$ bands. Their presence or absence from the spectra relates
to the degree of conformal symmetry realization in the light flavor sector.
Our next result is that the gauge interaction modifies the $S^3$ 
centrifugal barrier and removes the hydrogen like
degeneracies in the levels of the conformal band.
This contrasts the case of the linear in the energy 
stationary Schr\"odinger equation, which keeps respecting in the 
interacting case the degeneracies of the ${\mathcal K}^2$ eigenvalue problem
\cite{raposo}.  
We furthermore observed that with the increase of the excitation energies,
when  the influence of the gauge potential gradually fades away,
and the spectrum approaches that of the free conformal rigid rotor, 
the data fit becomes better. Above observations clearly point toward 
relevance of conformal symmetry for the spectra of the light flavor baryons.  
Conformal symmetry in the $N$ and $\Delta$ spectra is not an exact symmetry.
The model presented accounts for this circumstance through the modification 
of the centrifugal barrier of the
conformally invariant free geodesic motion on $S^3$ through the gauge
interaction. This modification is responsible for the systematic
$P_{2I,1}$--$S_{2I,1}$ mass splitting, which finds a 
satisfactory explanation within the framework under discussion.
Encouraging, the reasonable shape of the instantaneous 
effective gluon propagator obtained  as a Fourier transform of 
the gauge potential.

\noindent
Work supported by CONACyT-M\'{e}xico under grant number
CB-2006-01/61286. Joint contribution based on  
talks by both of us at the XII Mexican Workshop on Particles and Fields.

\end{document}